\documentclass[aps,prb,reprint]{revtex4-2}
\usepackage[utf8]{inputenc}
\usepackage{graphicx} % Required for inserting images
\usepackage{mathtools}
\usepackage{amsmath,amssymb,bm}
\usepackage{geometry}
\usepackage{multirow}
\begin{document}

\title{Spin-1 Dirac dispersion and Chern insulating phases in 2D honeycomb Sierpi\'nski fractal}

\author{Shneha Biswas$^{1}$, Shouya Yoshida$^{2}$, Katsunori Wakabayashi$^{2,3}$ and Sudipta Dutta$^{1}$}
\affiliation{$^{1}$Department of Physics, Indian Institute of Science Education and Research (IISER) Tirupati, Tirupati - 517619, Andhra Pradesh, India \\
$^{2}$Department of Nanotechnology for Sustainable Energy, School of Science and Technology, Kwansei Gakuin University, Gakuen-Uegahara 1, Sanda 669-1330, Hyogo, Japan \\
$^{3}$Research Center for Materials Nanoarchitectonics (MANA), National Institute for Materials Science (NIMS), 1-1 Namiki, Tsukuba, Ibaraki 305-0044, Japan
}

% Uncomment if multiple authors
% \author{Coauthor Name}
% \affiliation{Coauthor Institute}

%\section{abstract}
\begin{abstract}
	Graphene-based Sierpi\'nski fractals host a zero-energy chiral mode and spin-1 Dirac dispersions within the nearest-neighbor tight-binding model. However, the presence of complex next-nearest neighbor hopping arising from the local flux and the staggered Semenoff mass terms, modeled within the Haldane Hamiltonian, breaks the time-reversal and spatial inversion symmetries, respectively, and makes these flat bands dispersive. Moreover, they introduce rich topological phases in this class of systems that can be characterized by Chern numbers up to $\pm 3$, i.e., beyond the conventional honeycomb lattice. These observations pave the way for the exploration of 2D periodic fractals beyond 
    graphene, where topological phase transitions can be realized through externally applied fields.
\end{abstract}

\maketitle
\section{Introduction}
In recent years, the study of topological phases of matter has become a central frontier in condensed-matter physics, driven by the discovery of topological insulators \cite{kane2005quantum,konig2007quantum,moore2007topological} and topological superconductors \cite{read2000paired,kitaev2001unpaired}. This field was initiated by the discovery of the quantum Hall effect by von Klitzing in 1980 \cite{klitzing1980new,laughlin1983anomalous,thouless1982quantized}. Topological phases are characterized by topological invariants \cite{hasan2010colloquium}, which are protected by the bulk energy gap and remain unchanged under continuous deformations of the electronic Hamiltonian that preserve the gap. For example, the Chern number, $\mathcal{C}$ \cite{thouless1982quantized,kohmoto1985topological} is a topological invariant in systems where time-reversal symmetry is broken, and it characterizes the band topology of the system. The Haldane model on the honeycomb lattice exhibits phases with $\mathcal{C}=\pm$1, corresponding to a single chiral edge mode and realizing the anomalous quantum Hall effect \cite{haldane1988model}. Phases with higher $\mathcal{C} (\ge$2) have been reported in extended Haldane model with third-nearest-neighbor hopping \cite{bhattacharya2017quenching}, in multilayer systems \cite{zhao2020tuning}, and in the Bishamon–kiko lattice \cite{ikegami2024topological}. Similarly, quantum spin Hall insulators, introduced by Kane and Mele \cite{kane2005quantum,kane2005z}, are characterized by a $\mathbb{Z}_{2}$ topological invariant.

Topological states have been extensively explored in a variety of lattice geometries, including honeycomb \cite{haldane1988model,kane2005quantum}, square \cite{hofstadter1976energy,stanescu2010topological,wu2016antiferromagnetic}, kagome \cite{ohgushi2000spin,guo2009topological}, and Lieb lattices \cite{weeks2010topological,goldman2011topological}. These systems fall within the tenfold-way classification scheme, which organizes topological phases according to the presence or absence of fundamental spectral symmetries, such as time-reversal, particle–hole, and chiral symmetries \cite{ryu2010topological}. However the tenfold-way accounts only for the integer dimensions. This limitation motivates the exploration of topological phases on non-integer dimensions. 

Fractals are self-similar structures at any length scale and are characterized by noninteger dimensions \cite{mandelbrot1982thefractalgeometryofnature,brune1994mechanism}. Prototypical examples include the Sierpi\'nski gasket and the Sierpi\'nski carpet. The interplay between noninteger dimensionality and quantum topology opens new routes toward unconventional topological phases that fall outside the conventional tenfold-way framework \cite{altland1997nonstandard,ryu2010topological}.  Brezezi\'nska et al. have investigated the quantum Hall effect on the Sierpi\'nski gasket \cite{brzezinska2018topology},  the first discussion of the fractional quantum Hall effect was presented, and the existence of anyons in these fractals was predicted \cite{manna2020anyons}. Superconductivity, higher order topology, and non-hermitian topology are also investigated on the Sierpi\'nski gasket \cite{manna2022higher,manna2024noncrystalline,manna2023inner}. Experimentally, the topological properties, higher-order topologies on fractals, are also explored \cite{yang2020photonic,biesenthal2022fractal,li2022higher,zheng2022observation,lai2024spin}, and the quantum spin Hall effect has been observed on a bismuth fractal on InSb \cite{canyellas2024topological}.

 So far, numerous theoretical studies on fractal topology have focused on quantum dots with predominant quantum phenomena \cite{osseweijer2024haldane,brzezinska2018topology}. Recent experiments have demonstrated the feasibility of synthesizing Sierpi\'nski-type fractal networks on substrates via molecular self-assembly and coordination chemistry \cite{cote2005porous,mo2019surface}. Depending on the intermolecular interactions, fractal structures can be stabilized by hydrogen/halogen bonding, metal–ligand coordination, or covalent linkages, with the later two offering significantly improved thermal stability \cite{wang2019construction}. Substrate templating further enables the formation of ordered arrays and crystalline fractal networks \cite{sarkar2014one,newkome2006nanoassembly}. Despite of experimental challenges in achieving controlled growth of large-area, high-generation fractals, theoretical exploration of topology in such non-Euclidean geometries are motivated by a few experimental successes and the huge potential of such systems in understanding new physics.

 In this paper, we investigate the fundamental topological properties of generation-1-Graphene-based fractal (gen-1-G-fractal), which is periodic in both $x$ and $y$ axes. The gen-1-G-fractal has $C_{3}$ rotational symmetry as shown below. The Haldane model on fractal geometries has been actively investigated, including studies on the Sierpi\'nski gasket and on two Sierpi\'nski gaskets connected along one edge \cite{canyellas2024topological,osseweijer2024haldane}. We study the Haldane model on this gen-1-G-fractal system and explore the topological phase diagrams for different filling fractions. In addition to such topological properties, flat bands provide another powerful route towards strongly correlated and topologically nontrivial phases. Over the past decade, flat bands in translationally invariant lattices have attracted intense interest due to their macroscopic degeneracy and sensitivity to interactions and disorder \cite{goda2006inverse,nishino2003flat,endo2010tight,bodyfelt2014flatbands}. Interestingly, the gen-1-G-fractal also hosts multiple flat bands. In this work, we therefore explore the interplay between fractal geometry, flat-band physics, and topological phases within the Haldane model.

\begin{figure}[t]
\centering
\includegraphics[width=\columnwidth,
    height=9cm,
keepaspectratio]{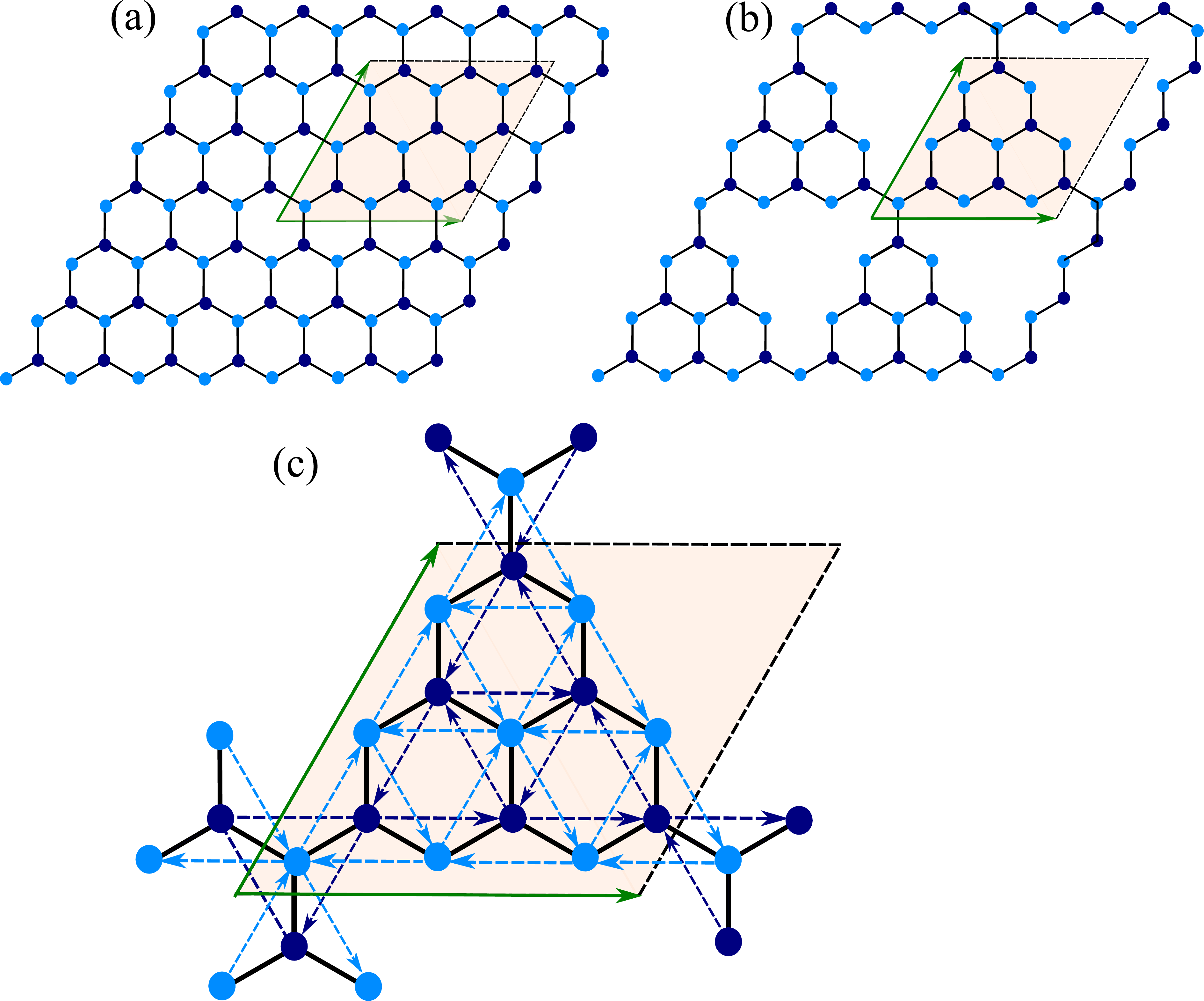}
\caption{Schematic of (a) graphene and (b) the generation-1-Graphene-based fractal. The two different shades of circles denote the two sublattices, A and B. The shaded rhombus indicates the unit cell of the fractal lattice, and the arrows represent the primitive lattice vectors. (c) Schematic of the Haldane model showing next-nearest-neighbor hoppings on the two sublattices with opposite circulation, as shown by the dashed arrows in one of the hexagonal plaquates.}
\label{fig:myfig}
\end{figure}

\section{ The Model}
We construct the first-generation-Graphene-based quantum fractal by periodically depleting 4 out of 18 lattice sites from the corner of a 3$\times$3 graphene supercell, as shown in Figs. 1(a) and 1(b). The resulting structure contains triangular pores with zigzag edges. Higher-order generations of this graphene-based fractal are illustrated in the Supplemental Material (SM). For our calculation, we focus on a first-generation-Graphene fractal (gen-1-G-fractal) periodically repeated over the two-dimensional (2D) plane. The corresponding fractal unit cell consists of 14 atoms, as shown in Fig. 1(b). The sublattice points are distinguished by two distinct shades, and the A- and B-sublattice sites within the fractal unit cell are unequal $N_{A}-N_{B}$ = 2 for the unit cell shown in Fig. 1(b). Focusing on the topological properties of the gen-1-G-fractal, we employ a spinless Haldane-like tight-binding model \cite{haldane1988model}, as follows.

\begin{figure}[t]
\centering
\includegraphics[width=\columnwidth,
    height=9cm,
keepaspectratio]{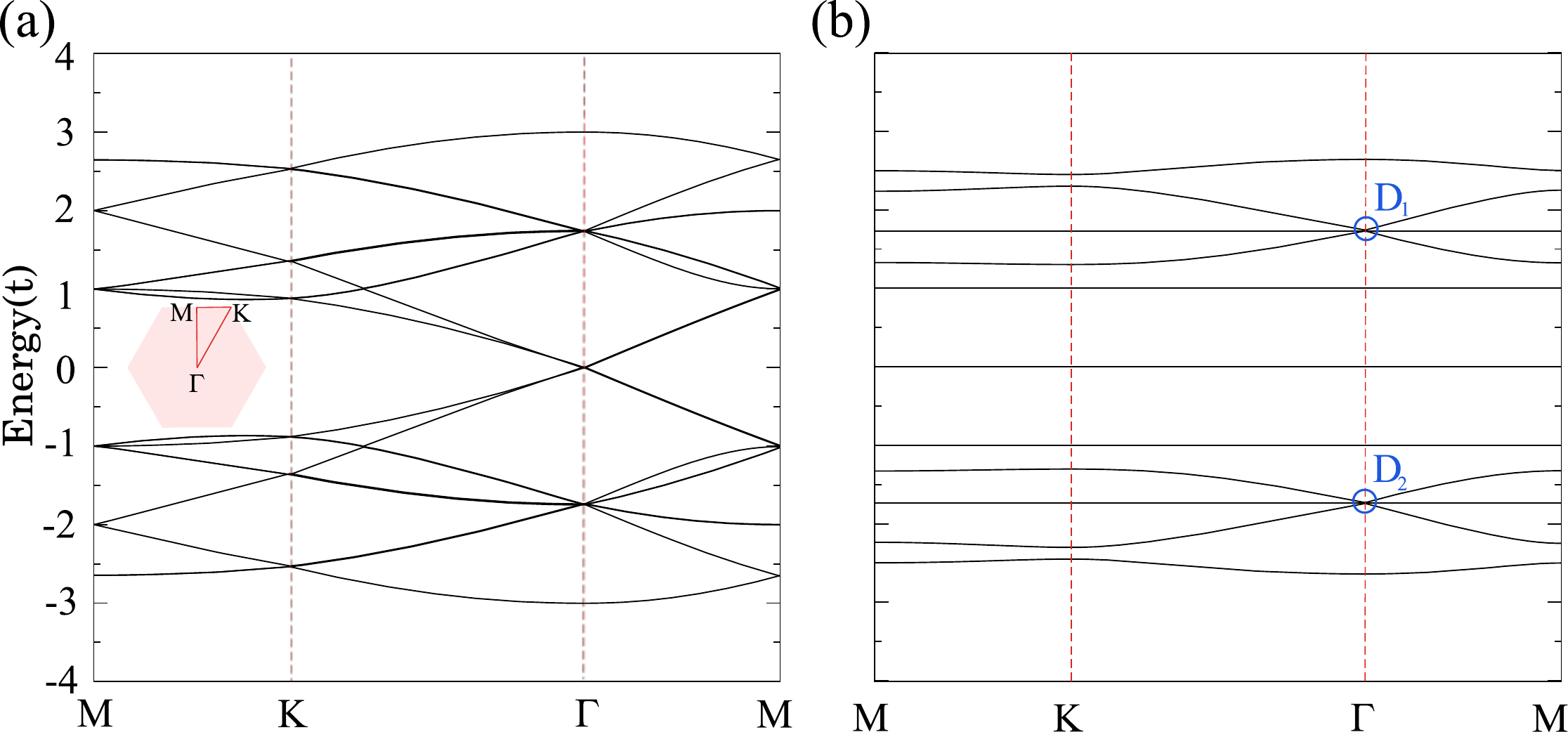}
\caption{Band dispersions of (a) honeycomb lattice with 3$\times$3 supercell (b) the generation-1-Graphene-based fractal modeled within nearest-neighbor tight-binding model, i.e. the Haldane Hamiltonian with \textbf{M} = 0 and $\lambda$ = 0. The Fermi energy is set to zero. The high-symmetry path is shown in the inset of (a) and the high-symmetry points are marked with vertical dashed lines. The spin-1 Dirac dispersions are highlighted by circles and denoted as D$_{1}$ and D$_{2}$ in (b).}
\label{fig:myfig}
\end{figure}

\begin{equation}
    H=t\sum_{<i,j>}\hat{c}^{\dagger}_{i}\hat{c}_{j} + \lambda\sum_{<<i,j>>} \hat{c}^{\dagger}_{i}\hat{c}_{j} e^{-i\mu_{ij}\phi} +\textbf{M}\sum_{i}\tau_{i} \hat{c}^{\dagger}_{i}\hat{c}_{i}
\end{equation}

 The first term of the Hamiltonian has the nearest-neighbor hopping (NN) with the hopping amplitude $t$, which is set as the unit of energy throughout. The second term represents the next nearest neighbor (NNN) hopping of strength $\lambda$, as depicted in Fig. 1(c). To break the time-reversal symmetry ($\mathcal{T}$), a magnetic flux density is applied perpendicular to the 2D plane. This flux density makes the NNN hoping term complex with a local flux $\phi$ \cite{haldane1988model}. Here $\mu_{i,j}$ is $+$1($-$1), which stands for clockwise (anticlockwise) NNN hopping, and $\phi$ denotes the associated local flux. The last term is a staggered Semenoff-type mass term of strength \textbf{M} that breaks the sublattice symmetry, where $\tau_{i}=$ $+$1($-$1) denotes sublattice A (B) \cite{semenoff1984condensed}. 

 \section{Results}
  Since the fractal lattice is constructed by selectively removing atoms from a graphene supercell, we examine the influence of fractal geometry by comparing the band structures of a pristine 3$\times$3 graphene supercell and the corresponding fractal unit cell in the absence of NNN and Semenoff-type mass term, as shown in Figs. 2(a) and 2(b), respectively. Due to the supercell structure and consequent folding of the Brillouin zone (BZ) along the momentum axis, the Dirac cone is mapped onto the $\Gamma$ point of the high symmetry path. The energy spectra of the gen-1-G-fractal contain multiple completely flat and dispersive bands, as can be seen in Fig. 2(b); some of them are degenerate. The isolated flat bands (FB) at energies $ E_{FB} = -1, 0, +1$ are doubly degenerate, while the other two flat bands at $E_{FB} =-\sqrt{3}, \sqrt{3}$ are non-degenerate. In particular, the zero-energy flat band can be identified as a chiral flat band originating from a sublattice imbalance. This is well known that for a bipartite lattice with only nearest-neighbor hopping, if the number of sites in the majority sublattice $N_{A}$ exceeds that in the minority sublattice $N_{B}$, there exist at least $N=|N_{A}-N_{B}|$ zero-energy eigenstates localized entirely on the majority sublattice \cite{sutherland1986localization,read2017compactly}. These flat bands are protected by the chiral symmetry of the system. However, upon introducing the complex NNN hopping term and the staggered Semenoff mass term, the chiral symmetry is broken. As a result, the zero-energy flat band gets shifted or becomes dispersive, as illustrated in Fig. 3(a), (b), and (c), thereby losing its chiral symmetry.

\begin{figure}[t]
\centering
\includegraphics[width= 1.0\columnwidth,
    height=9cm,
keepaspectratio]{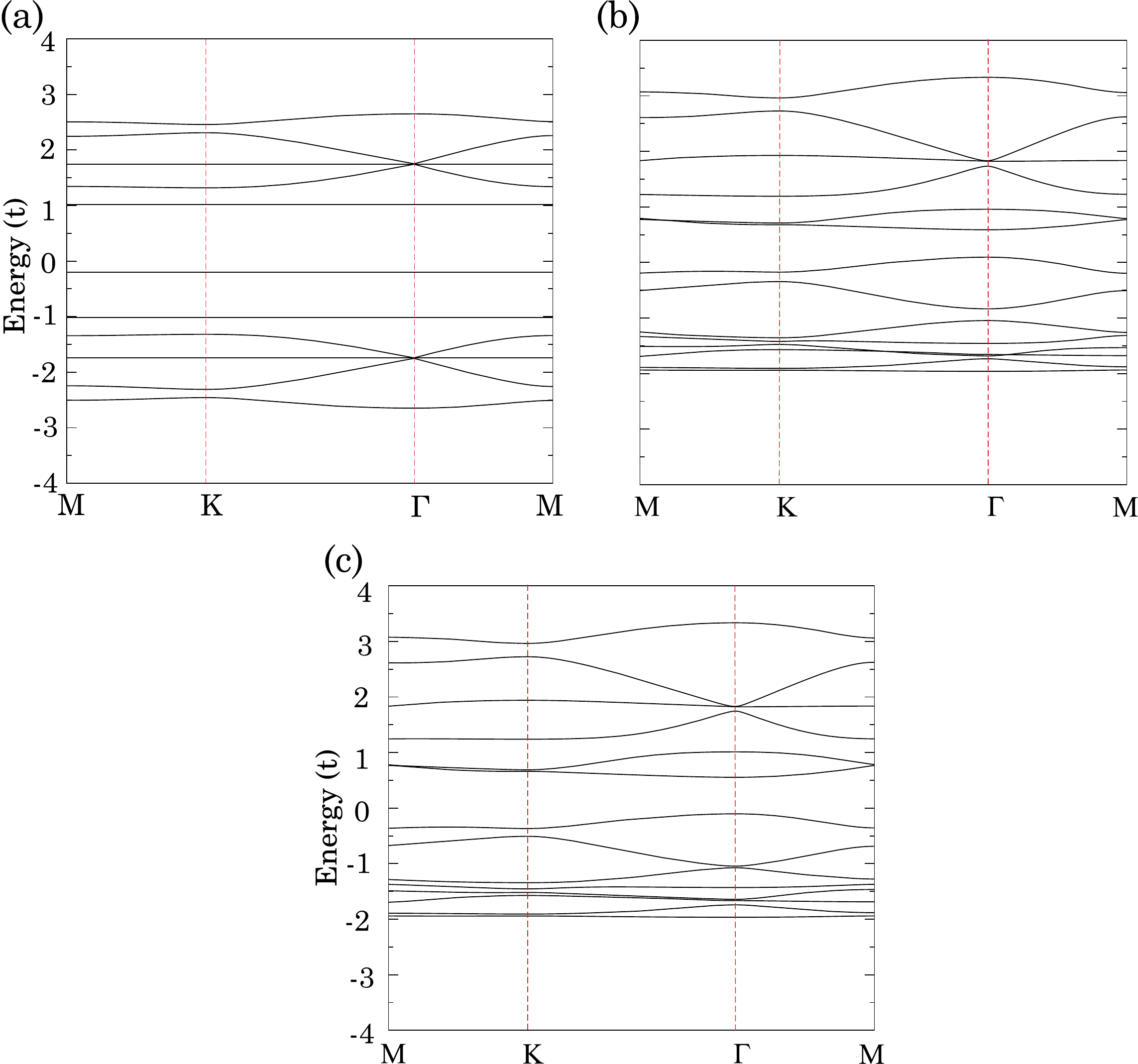}
\caption{Band dispersions of generation-1-Graphene-based fractal modelled within Haldane Hamiltonian with (a) \textbf{M} = 0.2, $\lambda$ = 0 (b) \textbf{M} = 0, $\lambda$ = 0.2, (c) \textbf{M} = 0.2, $\lambda$ = 0.2. The Fermi energy is set to zero, and the high-symmetry points are marked with vertical dashed lines. }
\label{fig:myfig}
\end{figure}

 Furthermore, the two non-degenerate flat bands at E $= \pm \sqrt{3}$ touch the Dirac cone formed by two dispersive bands at the $\Gamma$ point (D$_{1}$ and D$_{2}$ points in Fig. 2(b)), forming a spin-1 conical type spectrum. Such dispersion can also be seen in decorated graphene \cite{yan2021realization}, kagome lattice with staggered magnetic flux \cite{green2010isolated}, and Lieb lattice with zero flux \cite{vicencio2015observation}. These D$_1$ and D$_2$ points at $\mathbf{k}=0$ are special points because particles can have both zero and infinite effective masses. To examine the dispersion near the D$_1$, D$_2$ points, we derive the effective Hamiltonian near the energy $-\sqrt{3}$ and $+\sqrt{3}$. Setting $\mathbf{k}=(0,0)$ in the Hamiltonian, we evaluate the Hamiltonian at the $\Gamma$ point, where all Bl\"{o}ch phases reduce to unity. Diagonalizing this 14$\times$14 Hamiltonian reveals three-fold degenerate eigenvalues at E = $+\sqrt{3}$ and E = $-\sqrt{3}$, which define the low-energy subspace responsible for the D$_{1}$ and D$_{2}$ points, respectively. The corresponding eigenvectors are used to construct a unitary transformation into a symmetry-adapted basis.

 To describe the dispersion near $\Gamma$, we expand the Bl\"{o}ch Hamiltonian to linear order in momentum, retaining only the leading $O(\mathbf{k})$ terms. This corresponds to a $\mathbf{k.p}$-type expansion, where higher-order terms giving subleading corrections are neglected. Projecting the linearized Hamiltonian onto the low-energy subspace yields a 
3$\times$3 effective Hamiltonian governing the low-energy physics near the D$_1$ point, given by

\begin{widetext}
\begin{equation*}
    H_{eff}(\mathbf{k}) = \sqrt{3}I+
\begin{pmatrix}
\ 0 & \frac{1}{4}i\sqrt{3}k_{x}& \frac{1}{8}i(\sqrt{3}k_{x}-3k_{y}) \\
-\frac{1}{4}i\sqrt{3}k_{x} & 0& \frac{1}{8}i(\sqrt{3}k_{x}-3k_{y})\\
-\frac{1}{8}i(\sqrt{3}k_{x}-3k_{y})& -\frac{1}{8}i(\sqrt{3}k_{x}-3k_{y})& 0&
\end{pmatrix}+O(\mathbf{k}^{2})
\end{equation*}
\end{widetext}

 The energy dispersion near the D$_{1}$ point is given by 
%\begin{equation}
\begin {align}
    E_{k}^{\pm}&=\sqrt{3} \pm \frac{3\sqrt{k_{x}^2+k_{y}^2}}{4\sqrt{2}},\nonumber\\
    E_{k}^{0}&=\sqrt{3} 
\end{align}
A detailed derivation of the linearized Hamiltonian, the construction of the unitary transformation, and the projection onto the low-energy subspace are provided in the SM.

\begin{figure*}[t]
\centering
\includegraphics[width= 1.0\textwidth,
    height=15cm,
keepaspectratio]{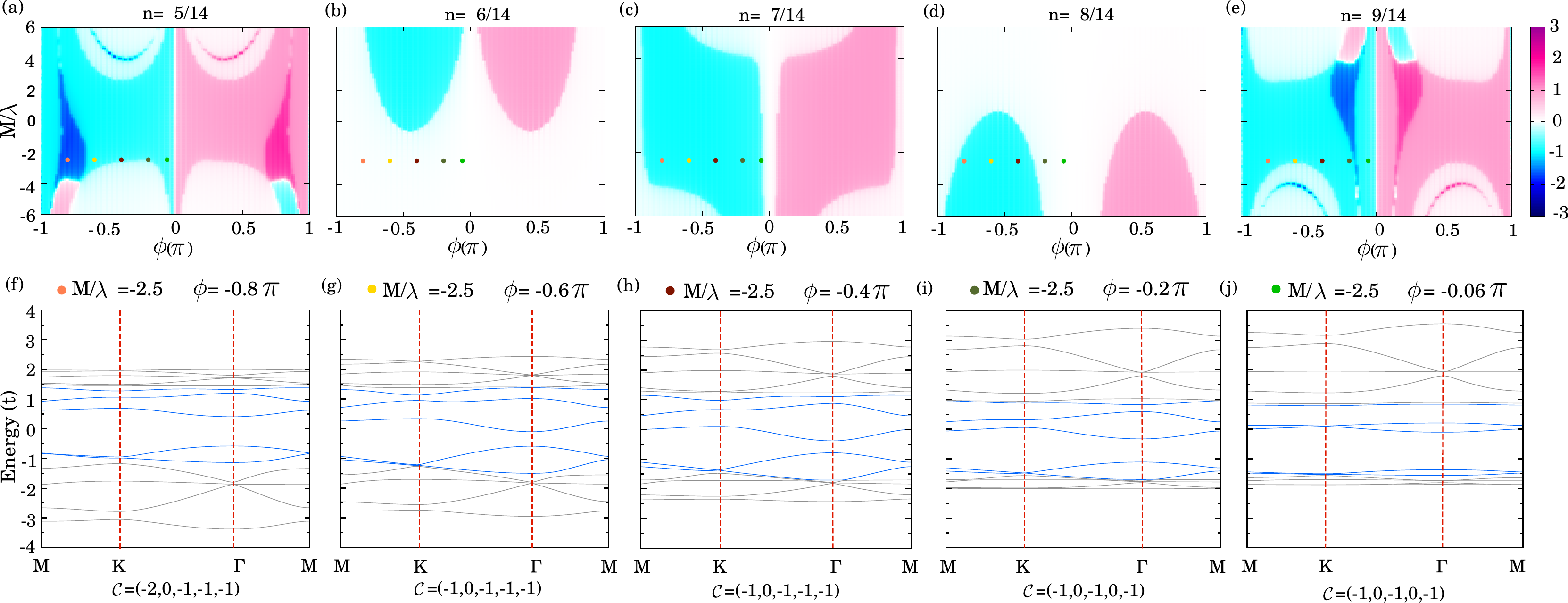}
\caption{Topological phase diagrams of the generation-1-Graphene-based fractal for band filling fractions, n$ =$  (a) $5/14$, (b) $6/14$, (c) $7/14$, (d) $8/14$, and (e) $9/14$. The color contour represents the Chern number. The band structures corresponding to the five marked points (colored circles in the phase diagrams) for $\phi$ = (f) $-$0.8$\pi$ (g) $-$0.6$\pi$ (h) $-$0.4$\pi$ (i) $-$0.2$\pi$ (j) $-$0.06$\pi$ at fixed \textbf{M}/$\lambda=-$2.5 value. The Chern numbers, $\mathcal{C}$, shown in each panel, correspond to the band fillings of $5/14$, $6/14$, $7/14$, $8/14$, and $9/14$, sequentially. The bands relevant to these fillings are highlighted in dark color.}
\label{fig:myfig}
\end{figure*}

However, the flat bands in the gen-1-G-fractal are highly fragile in the presence of complex NNN hopping and a staggered mass term, as implemented in the Haldane Hamiltonian. In the presence of only the staggered mass, the flat bands and spin-1 Dirac dispersions survive (see Fig. 3(a)). The presence of complex NNN hopping breaks the bipartite symmetry of the system, which makes the flat bands dispersive, lifts the degeneracy and destroys the spin-1 Dirac dispersions (see Figs. 3(b) and 3(c)).

 The complex NNN hopping term generates an effective staggered flux pattern, thereby breaking time-reversal symmetry and enabling the emergence of nontrivial topological phases. To characterize the topological nature of the bands, we compute the $\mathcal{C}$. Since the band structure exhibits degeneracies, the conventional Abelian Berry curvature formalism is not sufficient. Instead, we adopt the lattice gauge formulation proposed by Takahiro Fukui et al. \cite{fukui2005chern} which allows for a non-Abelian generalization of the Berry connection.

In this approach, the non-Abelian Berry connection is defined as \begin{equation}
    \textit{\emph{A}}=\psi^{\dagger}d\psi
\end{equation}
where $\psi=(|n_{1}\rangle,....,|n_{q}\rangle)$ denotes a multiplet of q occupied bands. The associated Chern number is 
\begin{equation}
    \mathcal{C}_{\psi}=\frac{1}{2\pi i}\int_{S}Tr dA
\end{equation}
where the integration is performed over the two-dimensional Brillouin zone surface $S$. On a discretized Brillouin zone, the link variable \textbf{U(1)} is defined by 
 \begin{equation}
     \textbf{U}_{\mu}(k_{l})=\frac{1}{N_{\mu}(k_{l})} \text{det} [\psi^{\dagger}(k_{l})\psi(k_{l}+\hat{\mu})] 
 \end{equation}
 where the normalization factor is defined as $N_{\mu}(k_{l})= |\text{det} [\psi^{\dagger}(k_{l})\psi(k_{l}+\hat{\mu})]|$ This formulation ensures gauge invariance and enables a stable numerical evaluation of the $\mathcal{C}$ even in the presence of band degeneracies. The Brillouin zone is divided into 5100 rhombi.

 To compare the topological phase transition as proposed by Haldane \cite{haldane1988model}, we plot the phase diagrams in Fig. 4 (a-e) as a function of \textbf{M}/$\lambda$ and local flux  $\phi$ at different filling factors. The color scale represents the corresponding $\mathcal{C}$. The $\mathcal{C}$ is evaluated only in regions where a finite band gap exists. Regions with non-zero integer values of $\mathcal{C}$ correspond to topological insulating phases, while the white regions with $\mathcal{C}=0$ represent trivial insulating phases. In the gapless regions, the $\mathcal{C}$ is ill-defined. The non-integer values appearing in the numerical phase diagrams are therefore used only as an indicator of gap closing and should not be interpreted as quantized topological invariants.
 
Due to the imbalance of the A and B sublattices, the phase diagrams are not symmetric with respect to the origin of \textbf{M}/$\lambda$. The phase diagram is symmetric between m/14 and 1$-$m/14 filling, where m = (1, 2,...13) by changing $\phi$ to $\phi+\pi$ and \textbf{M}/$\lambda$ to $-$\textbf{M}/$\lambda$. As can be seen, the phase diagrams corresponding to the filling factors of $5/14$ and $6/14$ are comparable with the phase diagrams corresponding to the filling factors of $9/14$ and $8/14$, respectively, through such symmetry. In the case of n = $5/14$ and $9/14$, we can observe two different types of topological phase transitions: (i) between two non-trivial topological phases characterized by integer Chern numbers ($|\mathcal{C}|\ge$1), irrespective of their sign, (ii) between the trivial insulating phase and topological insulating phases. However, in the case of n = $6/14$, $7/14$, and $8/14$, we observe only one type of phase transition between trivial to topological insulating phases. All such phase transitions happen through the metallic phase, as can be seen in Fig. 4(a - e).
 
For comparison, the conventional Haldane model on pristine graphene supports topological insulating phases with $\mathcal{C}=\pm$1. In periodically $1/6$-depleted (decorated) graphene, $\mathcal{C}=\pm$2 have been reported \cite{ikegami2024topological}. In contrast, the present gen-1-G-fractal exhibits even larger topological indices reaching $\mathcal{C}=\pm$3, demonstrating that fractal lattice engineering significantly enhances the topological diversity of the system.

Furthermore, upon increasing the strength of the NNN hopping $\lambda$, the topological phase diagram develops increasingly intricate structures, reminiscent of Arnold tongues \cite{adi_an1965eleven}. Similar tongue-like patterns have also been observed in fractal structures in the parameter space of dynamical systems \cite{boyland1986bifurcations}, Haldane model on  Sierpi\'nski gasket \cite{osseweijer2024haldane}, indicating a deep connection between lattice fractality and the emergence of complex topological phase boundaries. Increasing the values of $\lambda$, the sinusoidal behavior of the original Haldane model completely vanishes, and the fractality starts playing the interactive behavior. The detailed evolution of these Arnold–tongue–like features with increasing $\lambda$ is presented in the SM.

To understand the phase transition, we plot the band structures for different values of local flux $\phi$ at a fixed value of \textbf{M}/$\lambda$ of $-$2.5, as shown in Fig. 4(f - j). For comparison with the phase transition plots in Fig. 4(a - e) with different band fillings, band numbers 5 - 9 are highlighted in blue. In Fig. 4(f), the band structure is shown for $\phi$ = $-$0.8$\pi$ value. The topology of the insulating phases changes as we change the band filling. The corresponding Chern numbers, $\mathcal{C}$ = $-$2, 0, $-$1, $-$1, $-$1, are obtained for filling fractions of $5/14$, $6/14$, $7/14,$ $ 8/14,$ $ 9/14$, as can be seen from Fig. 4(a) to (e) at the specific values of $\phi$ and \textbf{M}/$\lambda$, depicted by orange circles. Some of these Chern numbers further vary with the $\phi$ values. As can be seen in Fig. 4 (f) and (g) for $5/14$ filling fraction, where up to the 5-th band is filled, the $\mathcal{C}$ changes from $-$2 to $-$1 via a band crossing. This indicates a topological phase transition through a metallic phase. A similar phase transition is observed at $8/14$ filling fraction, where the $\mathcal{C}$ changes from $-$1 to 0, as depicted in Fig. 4 (h) and (i). However, for filling fraction of $6/14$, the system remains in a trivial insulating phase at all $\phi$ values, owing to well-separated 6-th and 7-th bands. Variation of \textbf{M}/$\lambda$ value for a fixed $\phi$ can also induce the topological phase transitions. Phase diagrams corresponding to the other filling fractions are shown in the SM. All these topological phases can be accessed through experimentally tunable control knobs, namely $\phi$ and \textbf{M}. The local flux $\phi$ can be engineered in artificial lattice systems, such as photonic/acoustic lattices, cold-atom optical lattices, through laser-induced tunneling, and Floquet engineering, that generate synthetic gauge fields\cite{jaksch1998cold,goldman2016topological}. One can consider systems with broken spatial inversion symmetry, such as borocarbonitride systems \cite{adhikary2023valley, adhikary2024circular, biswas2025excitonic} or transition metal dichalcogenide systems with varying strengths of \textbf{M}. Furthermore, consideration of buckled 2D materials, such as silicene, allows the tuning of \textbf{M} through an external electric field \cite{ezawa2012topological}, that breaks the sublattice symmetry. Therefore, the present study on the gen-1-G-fractal can be extended to a plethora of other 2D materials with periodic fractal geometry.

\section{Conclusion}

To summarize, we investigate the 2D periodic fractal geometry of graphene, namely gen-1-G-fractal Sierpi\'nski gasket within the Haldane Hamiltonian with complex NNN hopping arising from local flux $\phi$ and with staggered Semenoff-type mass term \textbf{M}. The first term breaks the time-reversal symmetry, and the latter breaks the sublattice symmetry. In the absence of these two terms, the system exhibits a chiral flat band at zero energy, isolated gapped flat bands, and the formation of spin-1 Dirac cones that sandwich flat bands away from the Fermi energy. In the presence of a non-zero \textbf{M}, the chiral zero mode gets shifted, whereas the non-zero $\phi$ makes all these flat bands dispersive. Moreover, they introduce a variety of topological phases characterized by Chern numbers up to $\pm 3$. One can switch between these quantum phases by controlling the $\phi$ and \textbf{M} externally. Further explorations of different 2D periodic fractal systems may give rise to richer phase diagrams. In the present work, we focus on a noninteracting spinless fermionic model. It would be interesting to extend this study to spinful fermions and include many-body interactions, which may further enrich the interplay between fractal geometry, flat-band physics, and topology.

\begin{acknowledgments}
SB and SD acknowledge IISER Tirupati for Intramural Funding and SERB, Department of Science and Technology (DST), Govt. of India for research grant CRG/2021/001731. KW acknowledges the financial support from JSPS KAKENHI (Grants No. JP25K01609, No. JP22H05473), JST CREST (Grant No. JPMJCR19T1) and Basic Science Research Projects (Grant No. 2401203) from the Sumitomo Foundation. The authors acknowledge the Sakura Science Exchange Program, Japan Science and Technology Agency (JST), Govt. of Japan for finacial support for the collaboration.
\end{acknowledgments}

%\end{equation}
%\bibliographystyle{unsrt}
%\bibliographystyle{apsrev4-2}
%\bibliography{reference} % Replace 'references' with the name of your .bib file


% -----------------------------
%        Supplementary Info
% -----------------------------
\clearpage
%\onecolumngrid

\appendix

% Reset counters
\setcounter{figure}{0}
\setcounter{table}{0}
\setcounter{equation}{0}

% Redefine numbering
\renewcommand{\thefigure}{S\arabic{figure}}
\renewcommand{\thetable}{S\arabic{table}}
\renewcommand{\theequation}{S\arabic{equation}}

% Hyperref compatibility
%\renewcommand{\theHfigure}{S\arabic{figure}}
%\renewcommand{\theHtable}{S\arabic{table}}
%\renewcommand{\theHequation}{S\arabic{equation}}

\begin{titlepage}

\centering

\vspace*{3cm}

{\Huge \textbf{Supplemental Material} \par}

\vspace{1.5cm}

{\LARGE \textbf{Spin-1 Dirac dispersion and Chern insulating phases in 2D honeycomb Sierpi\'nski fractal} \par}

\vspace{2cm}

\author{Shneha Biswas$^{1}$, Shouya Yoshida$^{2}$, Katsunori Wakabayashi$^{2,3}$ and Sudipta Dutta$^{1}$}
\affiliation{$^{1}$Department of Physics, Indian Institute of Science Education and Research (IISER) Tirupati, Tirupati - 517619, Andhra Pradesh, India \\
$^{2}$Department of Nanotechnology for Sustainable Energy, School of Science and Technology, Kwansei Gakuin University, Gakuen-Uegahara 1, Sanda 669-1330, Hyogo, Japan \\
$^{3}$Research Center for Materials Nanoarchitectonics (MANA), National Institute for Materials Science (NIMS), 1-1 Namiki, Tsukuba, Ibaraki 305-0044, Japan
}

{\large
Shneha Biswas$^{1}$, Shouya Yoshida$^{2}$, Katsunori Wakabayashi$^{2,3}$ and Sudipta Dutta$^{1}$ \par
$^{1}$Department of Physics, Indian Institute of Science Education and Research (IISER) Tirupati, Tirupati - 517619, Andhra Pradesh, India \\
$^{2}$Department of Nanotechnology for Sustainable Energy, School of Science and Technology, Kwansei Gakuin University, Gakuen-Uegahara 1, Sanda 669-1330, Hyogo, Japan \\
$^{3}$Research Center for Materials Nanoarchitectonics (MANA), National Institute for Materials Science (NIMS), 1-1 Namiki, Tsukuba, Ibaraki 305-0044, Japan
}

\vfill

\end{titlepage}

\textbf{Effective Hamiltonian near the Triple degeneracy:}

We construct the tight-binding Hamiltonian for the generation-1-Graphene-based fractal in the absence of next-nearest-neighbor (NNN) hopping and staggered Semenoff-type mass. The eigenvalue equation is given by $\hat{H}(\mathbf{k})|u_{i}(\mathbf{k})\rangle=E_{i}|u_{i}(\mathbf{k})\rangle$ where $i=$1,..14 where $\textbf{k} = (k_{x}, k_{y})$. Instead of working with the full 14 $\times$ 14 Hamiltonian, we focus on the low-energy subspace at the triply degenerate point at E = $\pm \sqrt{3}$ at the $\Gamma$ point. Expanding the Bloch Hamiltonian up to first order in momentum, we write
$\hat{H}(\mathbf{k})=\hat{H}^{0}+\hat{H}^{1}+O(k^2),$ where $\hat{H}^{0}$ represents the Hamiltonian at the $\Gamma$ point and $\hat{H}^{1}$ contains the linear momentum-dependent terms. To obtain the effective low-energy description, we introduce a unitary transformation $U_{\Gamma}$, whose columns form an orthonormal basis of the threefold-degenerate eigenspace of the $\hat{H}^{0}$. Using this projection, the effective Hamiltonian around the triply degenerate point at $E=+\sqrt{3}$ is obtained as

\begin{widetext}
\begin{equation*}
    H_{eff} = U^\dagger_{\Gamma}\hat{H}^{1}_{\Gamma}U_{\Gamma}=\begin{pmatrix}
\ \sqrt{3} & 0&0 \\
0 & \sqrt{3}&0 \\
0& 0& \sqrt{3}&
\end{pmatrix}+
\begin{pmatrix}
\ 0 & \frac{1}{4}i\sqrt{3}k_{x}& \frac{1}{8}i(\sqrt{3}k_{x}-3k_{y}) \\
-\frac{1}{4}i\sqrt{3}k_{x} & 0& \frac{1}{8}i(\sqrt{3}k_{x}-3k_{y})\\
-\frac{1}{8}i(\sqrt{3}k_{x}-3k_{y})& -\frac{1}{8}i(\sqrt{3}k_{x}-3k_{y})& 0&
\end{pmatrix}+O(\mathbf{k}^{2})
\end{equation*}
\end{widetext}

Similarly, the effective Hamiltonian for the triply degenerate point at $E=-\sqrt{3}$ is given by

\begin{widetext}
\begin{equation*}
    H_{eff} = U^\dagger_{\Gamma}\hat{H}^{1}_{\Gamma}U_{\Gamma}=\begin{pmatrix}
\ -\sqrt{3} & 0&0 \\
0 & -\sqrt{3}&0 \\
0& 0& -\sqrt{3}&
\end{pmatrix}+
\begin{pmatrix}
\ 0 & -\frac{1}{4}i\sqrt{3}k_{x}& \frac{1}{8}i(\sqrt{3}k_{x}-3k_{y}) \\
\frac{1}{4}i\sqrt{3}k_{x} & 0& \frac{1}{8}i(\sqrt{3}k_{x}-3k_{y})\\
-\frac{1}{8}i(\sqrt{3}k_{x}-3k_{y})& -\frac{1}{8}i(\sqrt{3}k_{x}-3k_{y})& 0&
\end{pmatrix}+O(\mathbf{k}^{2})
\end{equation*}
\end{widetext}

These projected $3\times3$ Hamiltonians reveal the physics near the $\Gamma$ point at $E=\pm\sqrt{3}$, where threefold-degenerate point consists of one flat band and two linearly dispersing bands. The explicit representation of the orthonormal basis matrix $U_{\Gamma}$ at $E=+\sqrt{3}$ used for the projection is

\begin{widetext}
\begin{equation*}
    U_{\Gamma} =
\begin{pmatrix}
\ \frac{1}{2\sqrt{3}} & -\frac{1}{2\sqrt{3}} & \frac{\frac{2}{\sqrt{3}}-\sqrt{3}}{\sqrt{\frac{10}{3}+2(-\frac{2}{\sqrt{3}}+\sqrt{3})^{2}}} \\
0 & \frac{1}{2}&0 \\
0& -\frac{1}{2\sqrt{3}}& \frac{1}{\sqrt{3(\frac{10}{3}+2(-\frac{2}{\sqrt{3}}+\sqrt{3})^{2})}}&\\
0 & 0 & -\frac{1}{\sqrt{\frac{10}{3}+2(-\frac{2}{\sqrt{3}}+\sqrt{3})^{2}}}\\
\frac{1}{2\sqrt{3}}& 0 &\frac{1}{\sqrt{3(\frac{10}{3}+2(-\frac{2}{\sqrt{3}}+\sqrt{3})^{2})}}\\
-\frac{1}{2}& 0 & 0\\
-\frac{1}{2\sqrt{3}}&-\frac{1}{2\sqrt{3}}&0\\
\frac{1}{2}&0&0\\
-\frac{1}{2\sqrt{3}}&\frac{1}{2\sqrt{3}}&\frac{-\frac{2}{\sqrt{3}}+\sqrt{3}}{\sqrt{\frac{10}{3}+2(-\frac{2}{\sqrt{3}}+\sqrt{3})^{2}}}\\
0&-\frac{1}{2}&0\\
-\frac{1}{2\sqrt{3}}&0&-\frac{1}{\sqrt{3(\frac{10}{3}+2(-\frac{2}{\sqrt{3}}+\sqrt{3})^{2})}}\\
0&0&\frac{1}{\sqrt{\frac{10}{3}+2(-\frac{2}{\sqrt{3}}+\sqrt{3})^{2}}}\\
0&\frac{1}{2\sqrt{3}}&-\frac{1}{\sqrt{3(\frac{10}{3}+2(-\frac{2}{\sqrt{3}}+\sqrt{3})^{2})}}\\
\frac{1}{2\sqrt{3}}&\frac{1}{2\sqrt{3}}&0
\end{pmatrix}
\end{equation*}
\end{widetext}

\begin{widetext}
	
\begin{figure}[b]
    \centering
    \includegraphics[width=\textwidth]{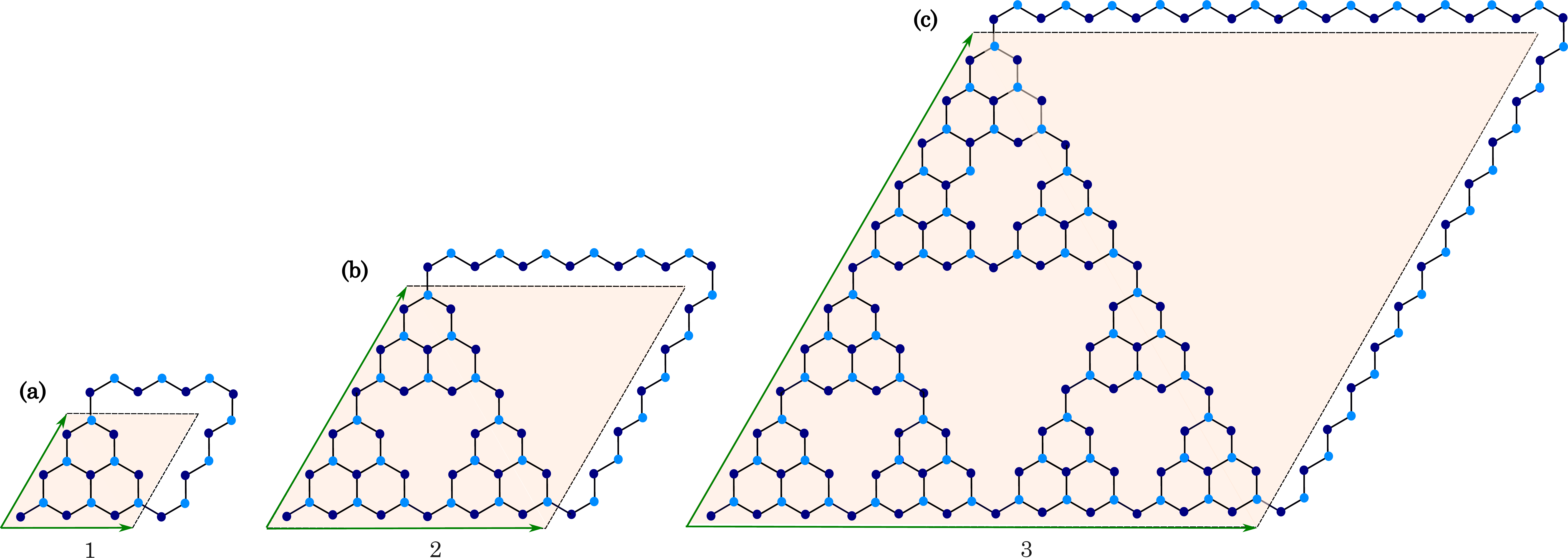}
    \caption{Graphene-based fractal structures of different generations (a) gen-1, (b) gen-2, (c) gen-3. The shaded region and green arrows denote the unit cell and primitive lattice vectors, respectively. }
\end{figure}

\begin{figure}[b]
    \centering
    \includegraphics[width=\textwidth]{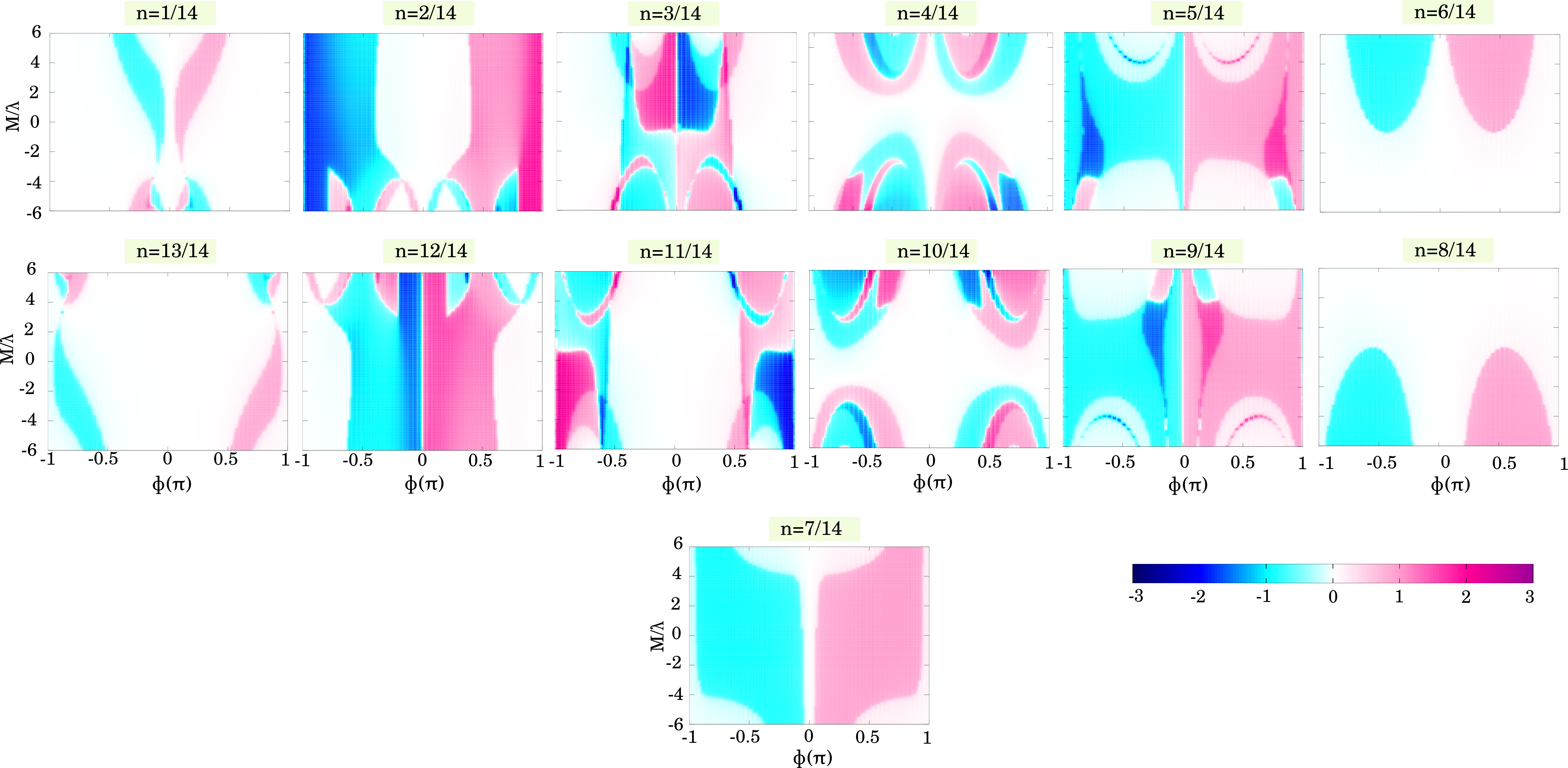}
    \caption{Topological phase diagrams of the Haldane model on the generation-1-Graphene-based fractal for different filling fractions (n) with $\lambda =$ 0.2. The color contour represents the Chern number. }
\end{figure}

\begin{figure}[b]
    \centering
    \includegraphics[width=\textwidth]{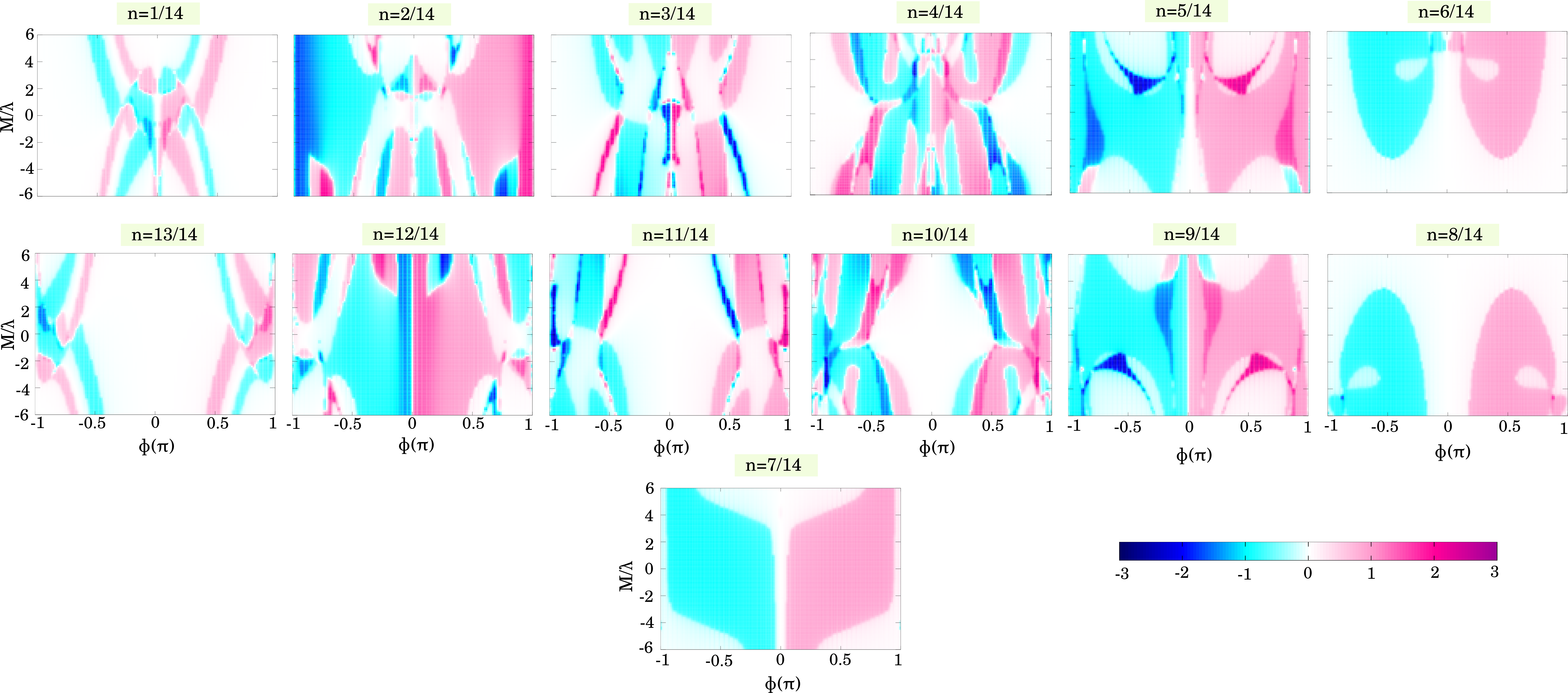}
    \caption{Topological phase diagrams of the Haldane model on the generation-1-Graphene-based fractal for different filling fractions (n) with $\lambda =$ 0.3. The color contour represents the Chern number. }
\end{figure}

\begin{figure}[b]
    \centering
    \includegraphics[width=\textwidth]{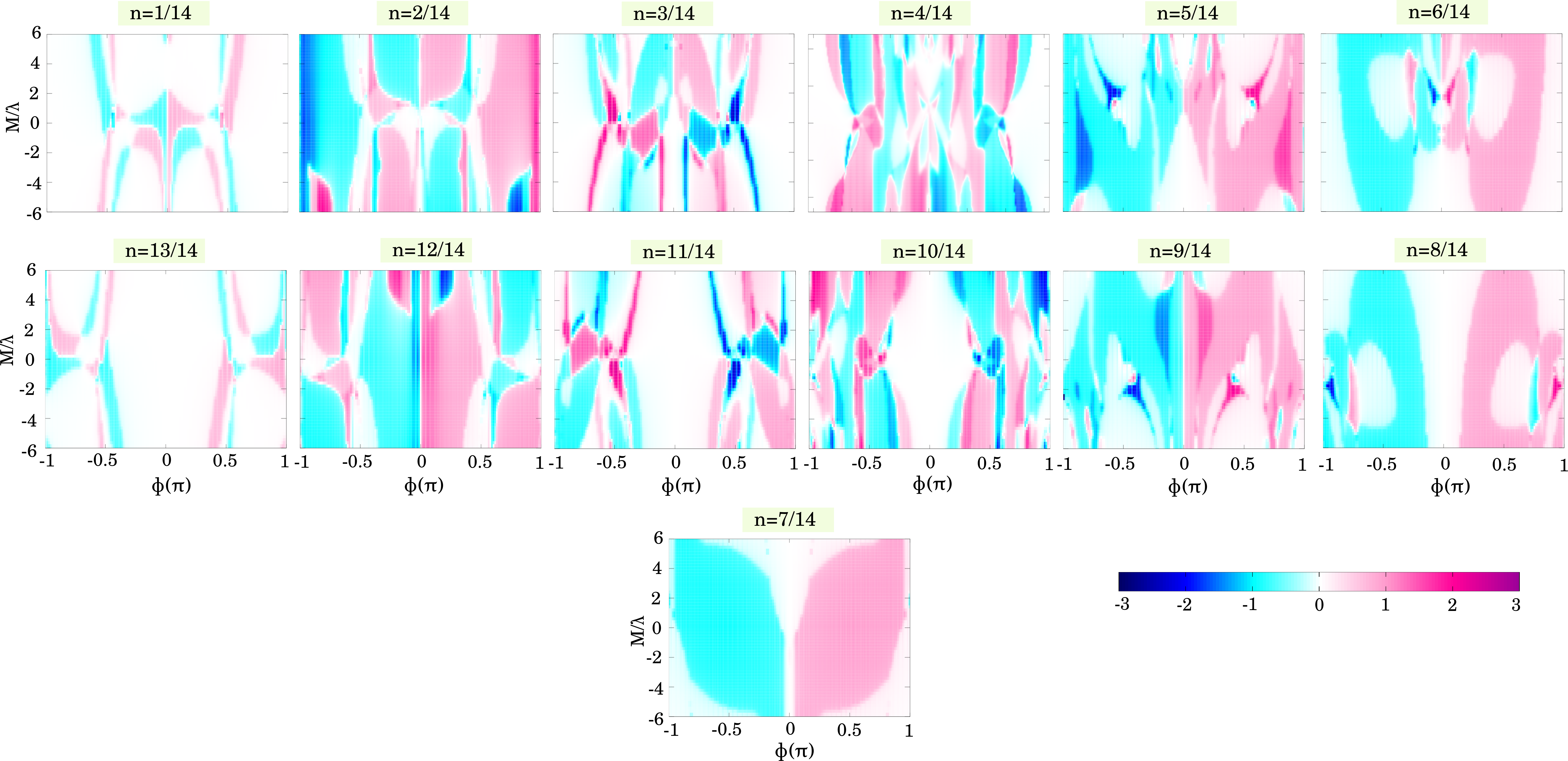}
    \caption{Topological phase diagrams of the Haldane model on the generation-1-Graphene-based fractal for different filling fractions (n) with $\lambda =$ 0.5. The color contour represents the Chern number. }
\end{figure}
\end{widetext}

\end{document}